\documentstyle[twocolumn,epsf,aps]{revtex}

\def\l{\langle}
\def\r{\rangle}

\begin{document}

\title{
Shape Effects of Finite-Size Scaling Functions for Anisotropic\\
Three-Dimensional Ising Models
}

\author{Kazuhisa Kaneda\cite{kaneda} and Yutaka Okabe\cite{okabe}}
\address{
Department of Physics, Tokyo Metropolitan University,
Hachioji, Tokyo 192-0397, Japan
}

\author{Macoto Kikuchi}
\address{
Department of Physics, Osaka University, Osaka 560-0043, Japan
}

\date{Received 22 July 1999}

\maketitle

\begin{abstract}
The finite-size scaling functions for anisotropic 
three-dimensional Ising models of size 
$L_1 \times L_1 \times aL_1$ ($a$: anisotropy parameter)
are studied by Monte Carlo simulations.
We study the $a$ dependence of finite-size scaling functions 
of the Binder parameter $g$ and the magnetization 
distribution function $p(m)$.
We have shown that the finite-size scaling functions 
for $p(m)$ at the critical temperature change from 
a two-peak structure to a single-peak one 
by increasing or decreasing $a$ from 1.
We also study the finite-size scaling near the critical 
temperature of the layered square-lattice Ising model, 
when the systems have a large two-dimensional anisotropy.
We have found the three-dimensional and two-dimensional 
finite-size scaling behavior depending on the parameter
which is fixed;  a unified view of 3D and 2D finite-size scaling 
behavior has been obtained for the anisotropic 3D Ising models.
\end{abstract}

\pacs{PACS numbers: 05.50+q, 75.10-b}

\vspace*{-3mm}

\section{Introduction}
Finite-size scaling (FSS) has been increasingly important 
in the study of critical phenomena \cite{fisher70,cardy88}.
This is partly due to the progress in the theoretical understanding 
of finite-size effects, and partly due to the application of 
FSS in the analysis of simulational results.

Recently, more attention has been paid to the universality of 
FSS functions \cite{pf84} for both percolation models 
\cite{hlc95a,hl96,lhc98,lh98} and Ising models \cite{ok96,wh97,okkh99}. 
It has been shown that several quantities including distribution 
functions on various lattices have universal 
FSS functions by choosing appropriate 
metric factors.  It should be noted that universal FSS functions 
depend on boundary conditions and the shape of finite systems.

In the percolation problem, it was considered until a quite 
recent time that there exists only one percolating cluster 
on two-dimensional (2D) lattices at the percolating threshold.  
But the importance of the number of percolating clusters for 
anisotropic systems, which was pointed out by Hu and Lin \cite{hl96}, 
has captured current interest \cite{hl96,aizenman97,sk97,cardy98}. 
Aizenman \cite{aizenman97} derived the upper and lower bounds of 
the probability for the appearance of $n$ percolating clusters, 
$W_n$, for 2D percolation at criticality.  His result was later 
confirmed by Monte Carlo simulations \cite{sk97}.  
By use of conformal field theory, Cardy \cite{cardy98}
proposed an exact formula for $W_n$ at the percolation threshold 
for the systems with large aspect ratios \cite{cardy98}. 
The aspect ratio $L_1/L_2$ is an important quantity in the 
FSS functions for anisotropic 2D systems of size $L_1 \times L_2$. 
The ``nonuniversal scaling'' of the 
low-temperature conductance peak heights for Corbino disks 
in the quantum Hall effect has been discussed 
in terms of the number of the percolating clusters \cite{chhr97}.

For the Ising model, the aspect ratio dependence of Binder 
parameter \cite{binder81} at the critical temperature was studied 
by Kamieniarz and Bl\"ote \cite{kb93}.  
Binder and Wang discussed anisotropic FSS \cite{binder89}.
The aspect ratio dependence of the universal FSS functions 
for Binder parameter and magnetization distribution function 
was discussed by Okabe et al. \cite{okkh99} for the 2D 
Ising model with tilted boundary conditions.  
For fixed set of the aspect ratio and the tilt parameter, 
the FSS functions were shown to be universal \cite{okkh99}. 
Quite recently, based on the connection between the Ising model 
and a correlated percolation model, Tomita et al. \cite{toh99} 
studied the FSS properties of distribution 
functions for the fraction of lattice sites in percolating 
clusters in subgraphs with $n$ percolating clusters, $f_n(c)$, 
and the distribution function for magnetization 
in subgraphs with $n$ percolating clusters, $p_n(m)$. 
They studied the change of the structure of 
the magnetization distribution function for the 2D system 
with large aspect ratio in terms of percolating clusters.

Almost all results above for the shape effects on FSS functions 
are for 2D systems except for some works on percolation problem
\cite{lhc98,lh98}.  It is quite important 
to extend these arguments for higher-dimensional systems. 
Conformal invariance plays a role in 2D systems \cite{cardy}, but is not 
so powerful for three-dimensional (3D) systems as for 2D ones.  
Moreover, there are two limiting cases for anisotropic 3D systems, 
that is, the one-dimensional (1D) limit and the 2D limit.
It is interesting to study various types of scaling for 
anisotropic 3D systems.

In this paper, we study the FSS functions for anisotropic 3D systems 
by Monte Carlo simulations.  We are concerned with 
the ferromagnetic Ising model on the $L_1 \times L_1 \times aL_1$
simple cubic lattices with the periodic boundary conditions, 
where $a$ is regarded as the anisotropy parameter.
Attention is mainly paid to the effect of anisotropy,
and we study the $a$ dependence of the FSS functions 
for quantities such as the Binder parameter and 
the magnetization distribution function.

At the critical temperature for the 3D Ising model,
we find that FSS functions for the magnetization distribution 
function change from a two-peak structure to a single-peak one, 
when the anisotropy parameter $a$ is varied from 1. 
This behavior is observed both for the systems with 2D anisotropy 
and for those with 1D anisotropy.
When a system has a large 2D anisotropy, we find that 
finite-size systems with fixed anisotropy parameter 
show good FSS behavior as 3D systems near 
the critical temperatures for the layered square-lattice Ising models.
In contrast, when we fix a number of layers and apply 
the FSS analysis for layered systems, these systems are 
scaled as 2D ones.

We organize the rest of the paper as follows.
In Sec.~2, we define the FSS functions and describe the 
quantities we treat in this paper. 
In Sec.~3, we present our simulational results.  
In Sec.~4, our results are summarized and discussed.

\section{Finite-size scaling}

If a quality $Q$ has a singularity of the form 
$Q(t) \sim t^\omega$ ($t=T-T_c$) near the criticality $t=0$,
then the corresponding quantity $Q(L,t)$ for finite systems
with the linear size $L$ has a following scaling form,
\begin{equation}
Q(L,t) \sim L^{-\omega/\nu} f(tL^{1/\nu}), 
\end{equation}
where $\nu$ is the correlation-length critical exponent
and $f(x)$ is the scaling function.  Of course, the corrections 
to FSS are not negligible for smaller $L$.
The FSS is also applicable to the distribution
function of $Q$. At the criticality $t=0$, the FSS 
function has a following form,
\begin{equation}
p(Q;L,t=0) \sim L^{\omega/\nu} F(Q L^{\omega/\nu}). 
\end{equation}

In this paper we focus on the Binder parameter and the 
magnetization distribution function.
The Binder parameter \cite{binder81} is given by 
\begin{equation}
g=\frac{1}{2} (3-\frac{\l m^4 \r}{\l m^2 \r^2}),
\end{equation}
and serves as a measure of the non-Gaussian nature of 
the distribution function. 
The FSS functions for these quantities are given by
\begin{eqnarray}
g(L,t) &\sim& g(tL^{1/\nu}),  \\
p(m;L,t=0) &\sim& L^{ \beta/\nu} p(mL^{ \beta/\nu}),
\end{eqnarray}
where $\beta$ is the magnetization exponent.

\section{Results}

We use the Metropolis Monte Carlo method 
to simulate the ferromagnetic Ising model on
$L_1 \times L_1 \times aL_1$ simple cubic lattices 
with different values of $L_1$ and $a$.
We use the periodic boundary conditions.

\begin{figure}
\centerline{\epsfxsize=\linewidth \epsfbox{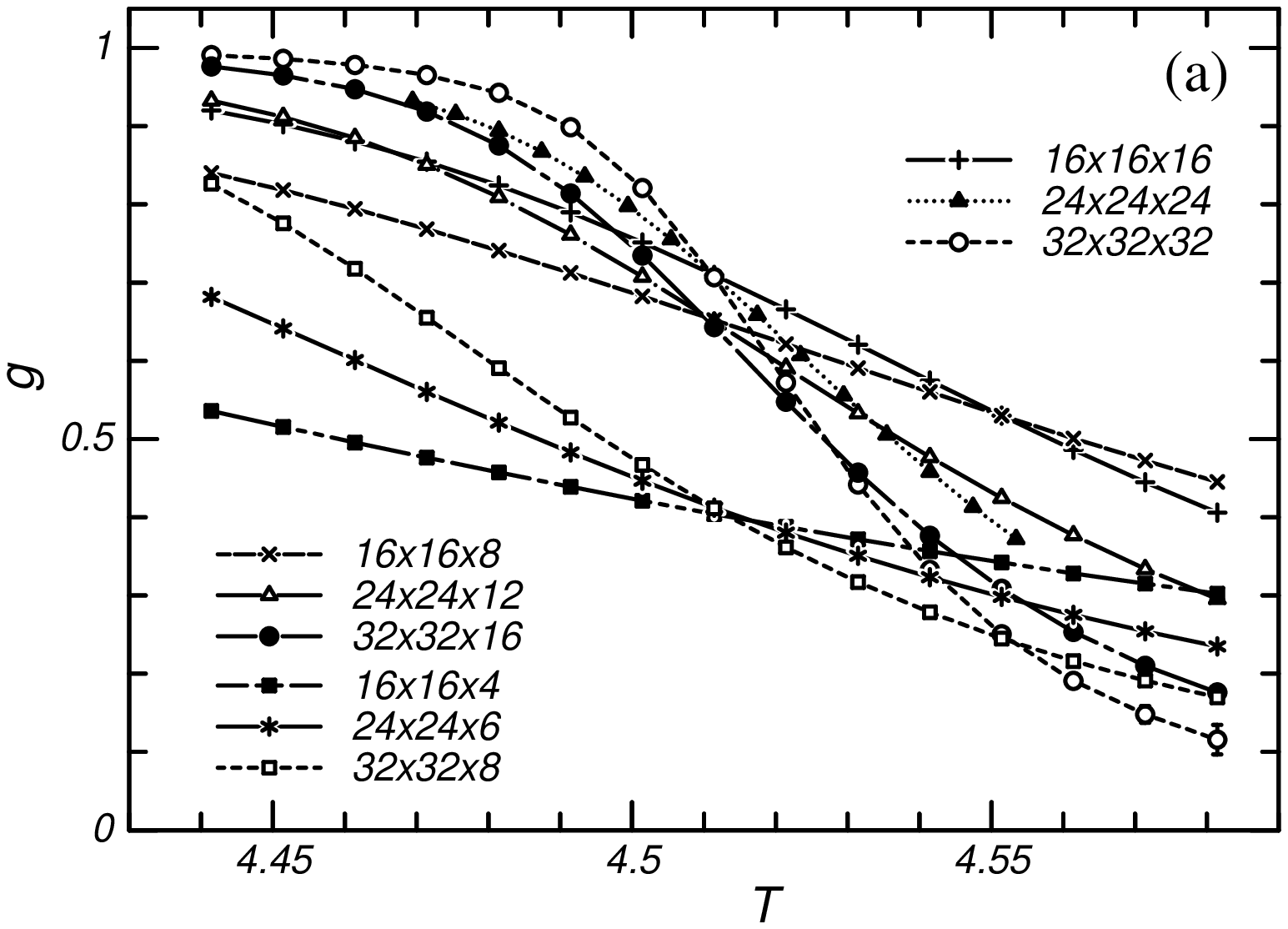}}
\vspace{4mm}
\centerline{\epsfxsize=\linewidth \epsfbox{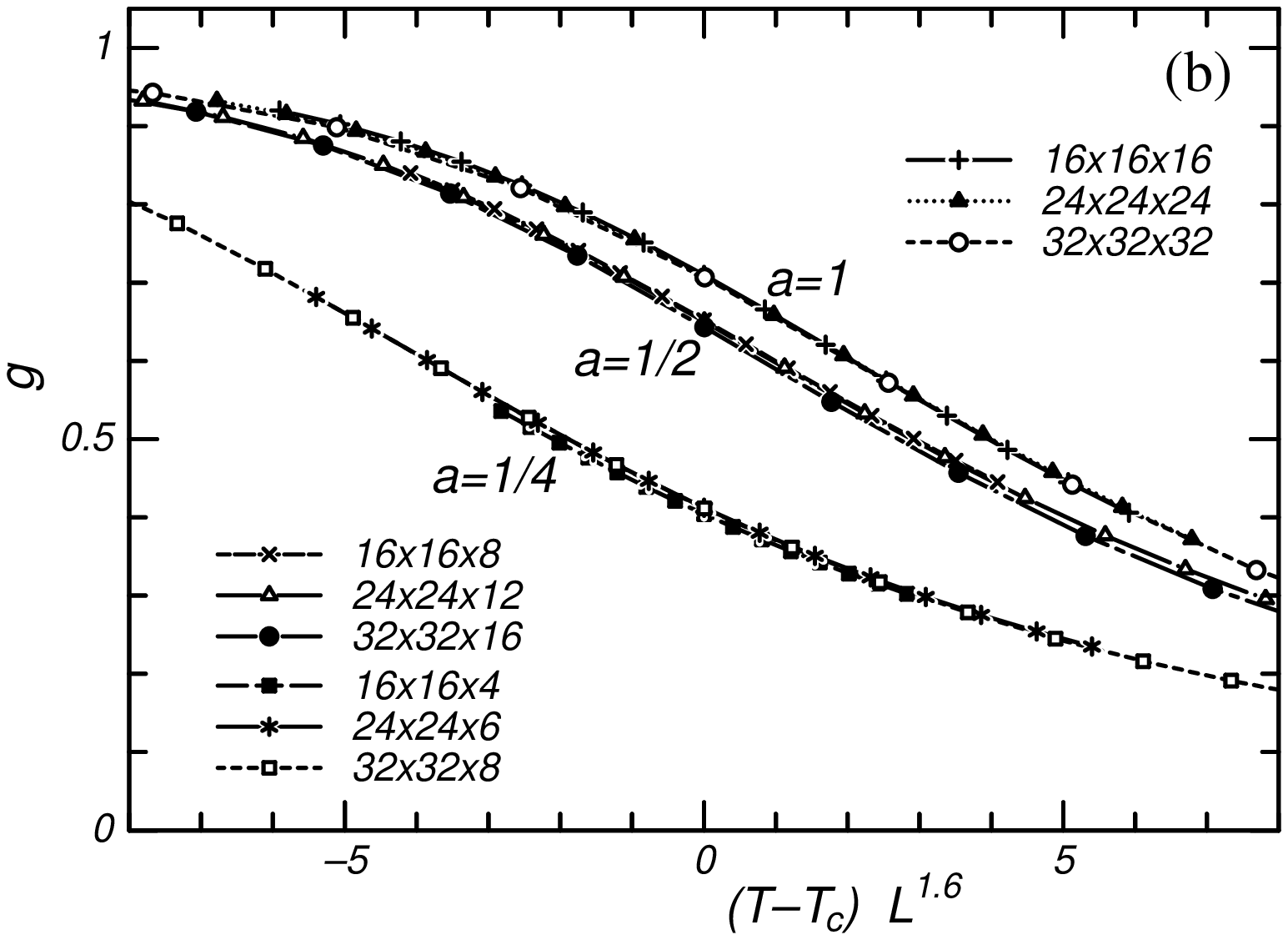}}
\vspace{4mm}
\caption{(a) Temperature dependence of $g$ for several lattices 
 with $a$=1, 1/2, and 1/4. The temperature is represented in unit of $J$.
(b) Plot of $g$ as a function of $(T-T_c) L^{1/\nu}$.}
\label{g_small}
\end{figure}
First, we show the results for $a \le 1$, that is,
a 2D anisotropic case.  We plot the temperature dependence 
of the Binder parameter $g$ for $a$=1, 1/2, and 1/4 of various 
sizes in Fig.~\ref{g_small}(a).  Error-bars are within 
the size of marks unless specified.  We give the FSS 
plots in Fig.~\ref{g_small}(b); 
$g$ is plotted as a function of $(T-T_c)L^{1/\nu}$, 
where $L$ is given by ($L_1 \times L_1 \times aL_1)^{1/3}$.
For $T_c$, $\beta$ and $\nu$, we use the numerically estimated 
values for the 3D Ising model \cite{fl91}, that is, 
$T_c=4.5114$, $\beta=0.320$ and $\nu=0.625$.
From now on, we represent the temperature in unit of $J$. 
From Fig.~\ref{g_small}(b) we see that the FSS functions 
have $a$ dependence in 3D systems, which is similar to 
the case of 2D systems.
\begin{figure}
\centerline{\epsfxsize=\linewidth \epsfbox{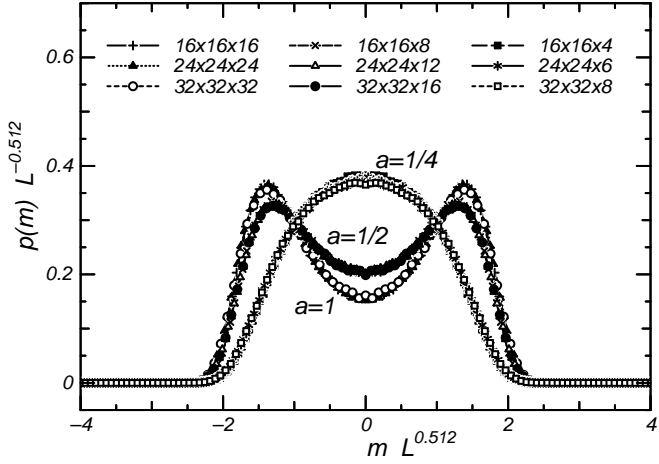}}
\vspace{4mm}
\caption{$p(m) L^{-\beta/\nu}$ at $T=T_c$ as a function 
 of $mL^{\beta/\nu}$ for several lattices with $a$=1, 1/2, and 1/4.}
\label{pm_small}
\end{figure}

We can get more information on the $a$ dependence from 
the magnetization distribution function $p(m)$.
In Fig.~\ref{pm_small}, we show the scaling plot of the magnetization 
distribution function $p(m)$ at $T=T_c$ 
for various sizes with different anisotropy parameters 
($a \le 1$).  From the figure we find good FSS behavior and also 
a large $a$ dependence.  By decreasing the anisotropy parameter $a$ 
from 1, the FSS functions for the magnetization distribution function
change from a two-peak structure to a single-peak one.

\begin{figure}
\centerline{\epsfxsize=\linewidth \epsfbox{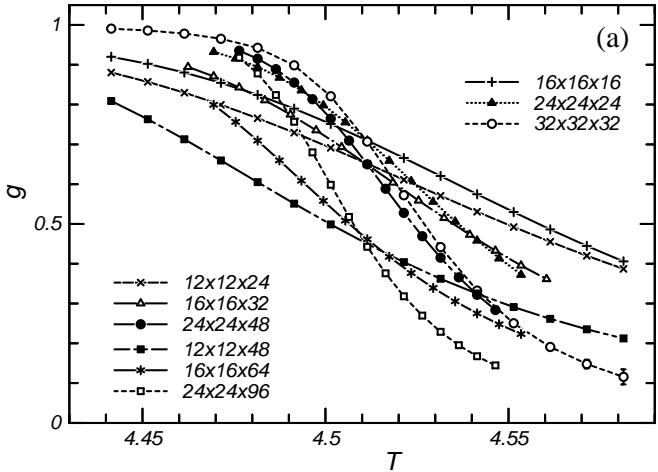}}
\vspace{4mm}
\centerline{\epsfxsize=\linewidth \epsfbox{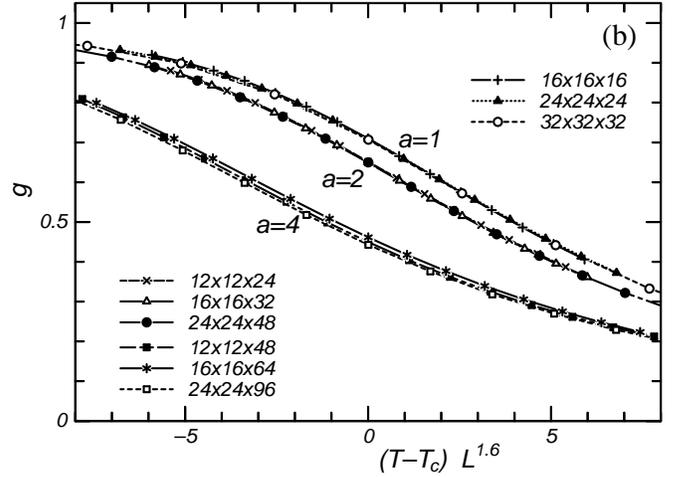}}
\vspace{4mm}
\caption{(a) Temperature dependence of $g$ for several lattices 
 with $a$=1, 2, and 4.
(b) Plot of $g$ as a function of $(T-T_c) L^{1/\nu}$.}
\label{g_large}
\end{figure}

\begin{figure}
\centerline{\epsfxsize=\linewidth \epsfbox{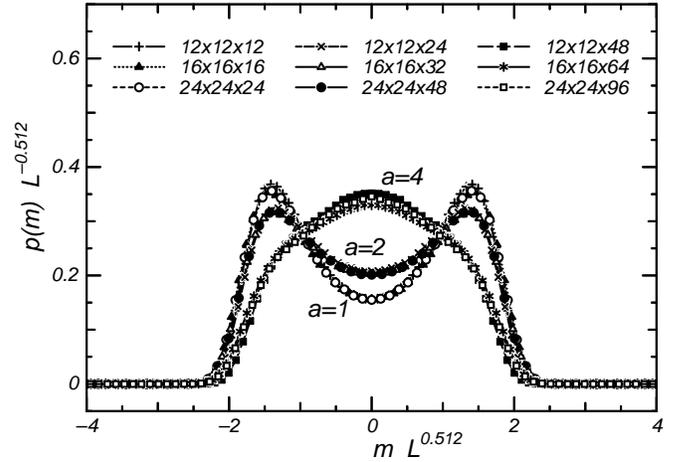}}
\vspace{4mm}
\caption{$p(m) L^{-\beta/\nu}$ at $T=T_c$ as a function 
 of $mL^{\beta/\nu}$ for several lattices with $a$=1, 2, and 4.}
\label{pm_large}
\end{figure}

Next turn to the case of $a \ge 1$, that is, a 1D anisotropic case.  
We plot the temperature dependence of $g$ for $a$=1, 2 and 4 
of various sizes in Fig.~\ref{g_large}(a).  The FSS plots are given 
in Fig.~\ref{g_large}(b).  
We see similar behavior of the FSS functions 
for $a \ge 1$, that is, 
$g$ take smaller values if we change $a$ from 1.
We show the $a$ dependence of the FSS functions of 
the magnetization distribution functions at $T=T_c$ 
for $a \ge 1$ in Fig.~\ref{pm_large}. 
By increasing $a$ from 1, FSS functions for $p(m)$ 
change from a two-peak structure to a single-peak one again.

If we compare the results of Figs.~\ref{g_small}(b) and 
\ref{g_large}(b), and also those of Figs.~\ref{pm_small} and 
\ref{pm_large}, it seems that there could be 
a set of the anisotropy parameters $a < 1$ and $a > 1$ 
which give the same FSS functions.  But it needs 
more careful calculations to clarify this universality.

Let us consider the origin of the $a$ dependence of FSS functions.
This problem is related to the multiple percolating clusters.
The connection between critical phenomena of spin models 
and percolation problems has been studied 
for a long time \cite{kf69,ck80,hu84}. 
Quite recently, Tomita et al. \cite{toh99} have
used the cluster formalism to investigate the percolating properties 
of the 2D Ising model.  They elucidated that the existence of 
several percolating clusters for anisotropic finite systems 
leads to the change of the structure of $p(m)$.
That is, the combination of up-spin clusters and down-spin clusters 
gives the contribution of $m \sim 0$ in $p(m)$ for anisotropic case.
The $a$ dependence of $p(m)$ for the 3D systems can be understood 
in the same way as the 2D systems.  We can apply this argument 
both cases for 2D anisotropy ($a < 1$) and 1D anisotropy ($a > 1$).  
The value of Binder parameter $g$ at $T=T_c$, of course, reflects
upon the structure of the distribution function $p(m)$.

\begin{figure}
\centerline{\epsfxsize=\linewidth \epsfbox{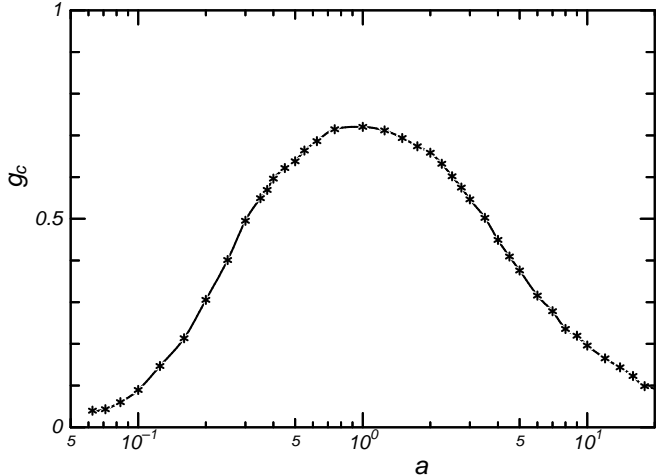}}
\vspace{4mm}
\caption{$g$ at $T=T_c$, $g_c$, as a function of $a$.  
We use a semi-logarithmic scale.}
\label{g_a_depend}
\end{figure}
In order to see the anisotropy parameter dependence more clearly, 
we plot the $a$ dependence of the Binder parameter $g$ 
at criticality, $g_c$, in Fig.~\ref{g_a_depend}.  Here, we use 
the semi-logarithmic scale for plot.  We add several data other than 
those given in Figs.~\ref{g_small} and \ref{g_large}.  
For large enough $a$ or small enough $a$, 
the corrections to FSS become larger.  
From Fig.~\ref{g_a_depend}, we find that $g_c$ takes
the maximum value at $a=1$ and decreases gradually 
in both directions, $a<1$ and $a>1$.  For very large $a$ 
or very small $a$, $g_c$ tends to vanish, which indicates 
that the distribution function $p(m)$ approaches 
the Gaussian distribution.  This behavior is consistent with 
above-mentioned multiple-percolating-cluster argument.
Such an $a$ dependence of $g$ 
at criticality for 1D anisotropy of 2D Ising model 
was already discussed for 2D Ising model \cite{kb93}.

\begin{figure}
\centerline{\epsfxsize=\linewidth \epsfbox{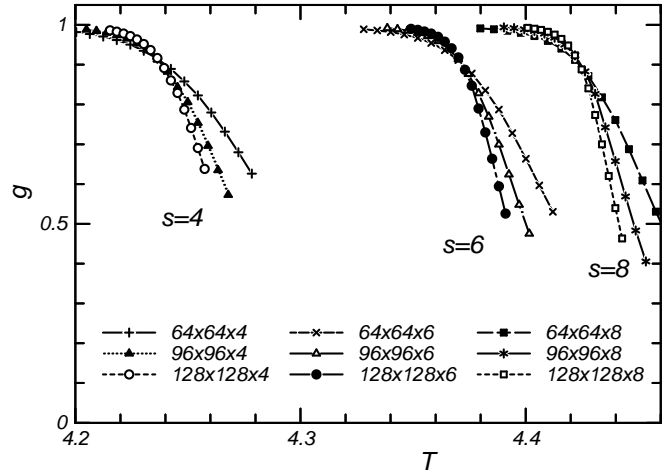}}
\vspace{4mm}
\caption{Temperature dependence of $g$ for several systems.
The numbers of layers are 4, 6, and 8.}
\label{g_layer}
\end{figure}

It is also interesting to consider the relation to 
the layered square-lattice Ising model for the case 
of large 2D anisotropy.  The critical temperatures 
and the shift exponent for the layered square-lattice
Ising model were studied by Kitatani et al. \cite{koi96}. 
In Fig.~\ref{g_layer}, we give the temperature dependence of 
$g$ for several lattices, where the number of layers, $s$, is 
4, 6, and 8.  If we make the system size large 
with fixing $s$, the system becomes the layered 
square-lattice Ising model.  We can estimate 
the $s$-dependent critical temperature $T_{cs}$ from 
the crossing point of the curves for fixed $s$ in Fig.~\ref{g_layer}.
Our estimates, $T_{c4}=4.2364(4)$, $T_{c6}=4.3701(4)$, and 
$T_{c8}=4.4222(4)$, are consistent with those of Ref.~\cite{koi96}.

\begin{figure}
\centerline{\epsfxsize=\linewidth \epsfbox{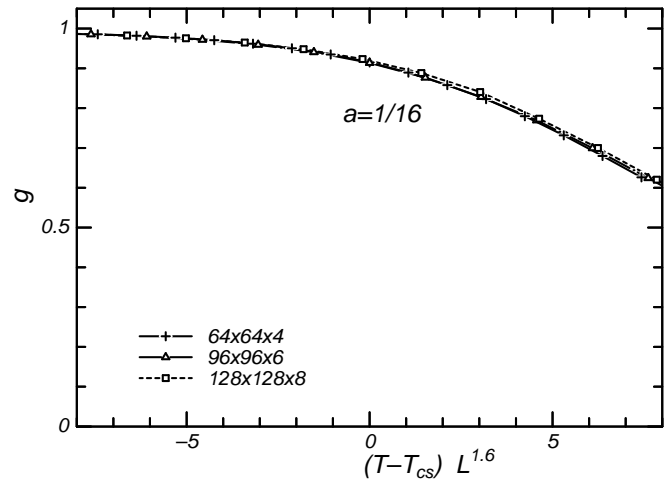}}
\vspace{4mm}
\caption{Scaling plot of $g$ for fixed $a (=1/16)$.  
The scaled variable is $(T-T_{cs}) L^{1/\nu}$ for each $s$.  
The values of $T_{cs}$ are $T_{c4}=4.2364$, $T_{c6}=4.3701$, 
and $T_{c8}=4.4222$. 
We use the 3D exponent, $\nu=0.625$.}
\label{g_3d}
\end{figure}

\begin{figure}
\centerline{\epsfxsize=\linewidth \epsfbox{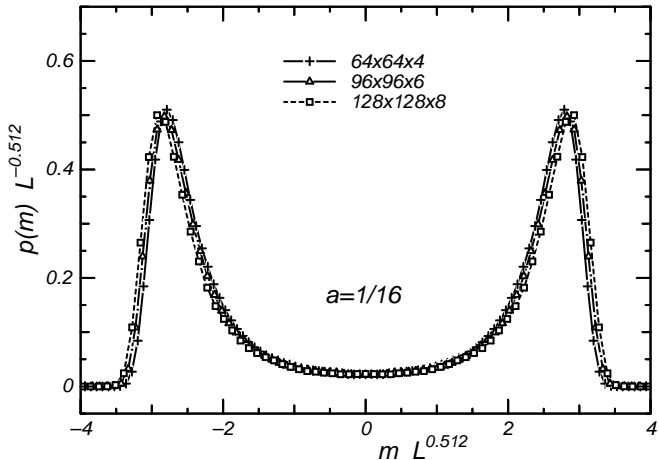}}
\vspace{4mm}
\caption{Scaling plot of $p(m)$ at $T=T_{cs}$ for 
fixed $a (=1/16)$; $p(m) L^{-\beta/\nu}$ as 
a function of $mL^{\beta/\nu}$. The values of $T_{cs}$ are 
$T_{c4}=4.2364$, $T_{c6}=4.3701$, and $T_{c8}=4.4222$. 
We use the 3D exponents, $\beta=0.320$ and $\nu=0.625$.}
\label{pm_3d}
\end{figure}

\begin{figure}
\centerline{\epsfxsize=\linewidth \epsfbox{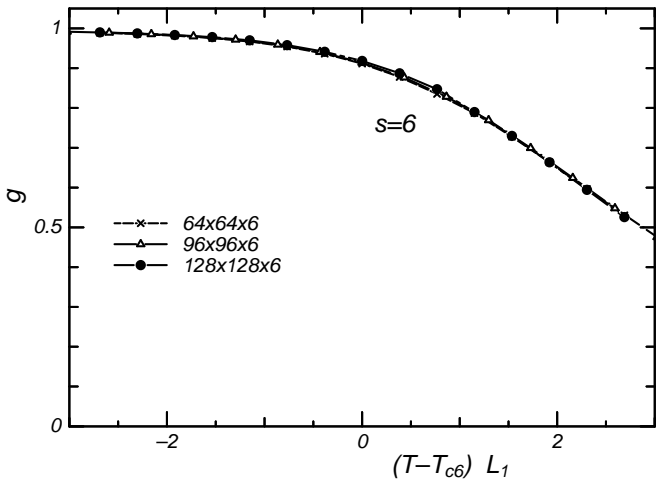}}
\vspace{4mm}
\caption{Scaling plot of $g$ for fixed $s (=6)$.  
The scaled variable is $(T-T_{c6}) L_1^{1/\nu}$, and 
$T_{c6}=4.3701$. We use the 2D exponent, $\nu=1$. }
\label{g_2d}
\end{figure}

Picking up the data for $a=1/16$ from Fig.~\ref{g_layer}, we plot 
$g$ as a function of $(T-T_{cs}) L^{1/\nu}$ in Fig.~\ref{g_3d}.
Here we treat $T-T_{cs}$ for each $s$, 
and $\nu$ is chosen as 0.625, that is, the 3D value.
We have good FSS behavior in Fig.~\ref{g_3d}, but this FSS plot is 
different from that given in Fig.~\ref{g_small}(b), where 
$T-T_c$ has been treated.  We should note that 
$T-T_{cs}$ can be rewritten as $T-T_{cs}=(T-T_c)+(T_c-T_{cs})$, and 
$T_c-T_{cs} \propto s^{-\lambda}$.  Here $\lambda$ is the shift exponent 
and is shown to be $1/\nu_{{\rm 3D}}$ \cite{koi96}.
Thus, for fixed $a$, we have 3D FSS for $T-T_{cs}$. 
In this case, the value of $g_c$ at $T=T_{cs}$ 
is $\sim 0.92$, which is nothing but the $g_c$ value for the isotropic 
2D Ising model.  It is different from $\sim 0.04$ for $g_c$ value of 
anisotropic 3D Ising model for $a=1/16$, which is given 
in Fig.~\ref{g_a_depend}.  In Fig.~\ref{pm_3d}, we also show 
the scaling plot of the magnetization 
distribution function $p(m)$ at $T=T_{cs}$ for each $s$, 
fixing $a$ as 1/16.  We have good FSS behavior using 3D exponents.  
There are two distinct peaks in $p(m)$, 
which is a typical character of $p(m)$ at $T_c$ 
for the isotropic 2D Ising model.
In contrast, if we pick up the data for fixed $s$, 
for example, $s=6$, these systems are expected to be 
scaled by using 2D critical exponents.  
Since we fix $s (= aL_1)$, the scaled variable is given by
$(T-T_{cs}) L_1^{1/\nu}$.
Using the 2D exponent, $\nu_{{\rm 2D}}=1$, we get very good 
scaling plot, shown in Fig.~\ref{g_2d}.  Of course, 
the critical value $g_c$ at $T=T_{c6}$ is $\sim 0.92$, that is,
the $g_c$ value for the isotropic 2D system.

\section{Summary and Discussions}
To summarize, we have studied the FSS functions for anisotropic 
3D Ising model by Monte Carlo simulations.  
The anisotropy parameter ($a$) dependence of 
the FSS functions for the Binder parameter and the magnetization 
distribution function has been investigated.  
We have shown that the magnetization distribution functions change
from two-peak structures to single-peak ones as $a$ 
increases or decreases from 1.
This change of distribution functions may be 
attributed to the combination of up-spin percolating clusters 
and down-spin percolating clusters for anisotropic systems, 
which is the same as the case of 2D systems \cite{toh99}. 
We should also note that the distribution function $p(m)$ 
approaches the Gaussian distribution for large anisotropy, 
which is another indication of the 
multiple-percolation-cluster argument.
It is interesting to calculate the probability for the appearance 
of $n$ percolating clusters, $W_n$, for the anisotropic 
3D Ising model and to make cluster analysis, which will be 
left to a future study.

Moreover, we have considered the relation to the layered 
square-lattice Ising models for large 2D anisotropy.  
We have obtained 3D FSS behavior near the critical temperature 
of the layered square-lattice Ising models, $T_{cs}$, for fixed $a$.
In contrast, when we fix the number of layers, we have 
shown 2D FSS behavior near $T_{cs}$.
Thus, we have obtained a unified view of 3D and 2D FSS behavior 
for the anisotropic 3D Ising models.

We have studied the shape effects of FSS functions for 
3D Ising models.  FSS functions also depend on boundary 
conditions.  Extension of the 2D results for various boundary 
conditions, for example, tilted boundary conditions \cite{okkh99},
to 3D systems will be interesting. 
This study is now in progress.

\section*{Acknowledgments}
We would like to thank C.-K. Hu and Y. Tomita for valuable discussions.
The computation in this work has been done using the facilities of 
the Supercomputer Center, Institute for Solid State Physics, 
University of Tokyo.  This work was supported by a Grant-in-Aid 
for Scientific Research from the Ministry of Education, Science, 
Sports and Culture, Japan.

\end{document}